\documentclass[structabstract]{aa}

\usepackage[utf8]{inputenc}

\usepackage{amsmath}

\usepackage{calc}

\usepackage{longtable}
\usepackage{booktabs}
\usepackage{array}

\usepackage{color}

\usepackage[pdftex]{graphicx}
\usepackage{subfigure}

\usepackage{txfonts}
\usepackage{ae}

\usepackage{natbib}
\bibpunct{(}{)}{;}{a}{}{,} 

\usepackage{units}

\usepackage{textcomp}
\usepackage{gensymb}

\begin{document}

\title{Morphometric analysis in gamma-ray astronomy\\
  using Minkowski functionals}
\subtitle{Source detection via structure quantification}

\author{D.~Göring\inst{\ref{ecap}}
  \thanks{\email{daniel.goering@physik.uni-erlangen.de}}
  \and
  M.~A.~Klatt\inst{\ref{theo1},\ref{ecap}}
  \thanks{\email{michael.klatt@physik.uni-erlangen.de}}
  \and
  C.~Stegmann\inst{\ref{desy}}
  \and
  K.~Mecke\inst{\ref{theo1},\ref{ecap}}
}

\titlerunning{Morphometric analysis in gamma-ray astronomy}
\authorrunning{D.~Göring, M.~A.~Klatt et al.}

\institute{Erlangen Centre for Astroparticle Physics,
  Universität Erlangen-Nürnberg,
  Erwin-Rommel-Str. 1,
  D 91058 Erlangen\label{ecap}
  \and
  Institut für Theoretische Physik,
  Universität Erlangen-Nürnberg,
  Staudtstr. 7,
  D 91058 Erlangen\label{theo1}
  \and
  DESY,
  Platanenallee 6,
  D 15738 Zeuthen\label{desy}
}

\date{Received 21 January 2013 / Accepted 18 April 2013}


\abstract
{} 
{ H.E.S.S. observes an increasing number of large \emph{extended}
  sources. A new technique based on the structure of the sky map is
  developed to account for these additional structures by comparing
  them with the common point source analysis.}
{ Minkowski functionals are powerful measures from integral
  geometry. They can be used to quantify the structure of the counts
  map, which is then compared with the expected structure of a pure
  Poisson background. Gamma-ray sources lead to significant deviations
  from the expected background structure.
  The standard likelihood ratio method is exclusively based on the
  number of excess counts and discards all further structure
  information of large extended sources. The morphometric data
  analysis incorporates this additional geometric information in an
  unbiased analysis, i.e., without the need of any prior knowledge
  about the source.}
{ We have successfully applied our method to data of the
  H.E.S.S. experiment. The morphometric analysis presented here is
  dedicated to detecting faint extended sources.}
{}

\keywords{
  Methods: data analysis --
  Methods: statistical --
  Techniques: image processing --
  Gamma rays: diffuse background
}

\maketitle


\section{Introduction}
\label{Introduction}

The aim of this study is to introduce a novel approach to data
analysis in very high energy (VHE) gamma-ray astronomy, where extended
sources are detected via morphometric valuations.

Early studies in VHE gamma-ray astronomy focused on the study of point
sources. For this purpose, highly efficient analysis techniques were
established, such as the quantification of the significance of a
photon count excess using a likelihood ratio method \citep{lima}. But
the increasing number of large extended sources
\citep[e.g.][]{rxj1713, velajr} and first detection of diffuse VHE
emissions \citep{galcen} emphasize the need for new approaches that
might be more suitable for extended structures.

An analysis based exclusively on the number of excess counts above the
expected background level discards all information about the shape of
the region where the excess is observed. Furthermore, it does not use
the information of possible correlations of nearby excess regions.
While this additional information is negligible for point sources, it
might provide a means to detect and study faint extended sources,
which are too weak to be seen when looking at the amount of excess
photons only.

A well-known approach to include all available information in an
analysis is the full likelihood fit of a model to the measured data,
as used by high-energy gamma-ray telescopes like EGRET
\citep{egretlike} or Fermi/LAT \citep{fermilat}. While likelihood
analyses are very powerful, they require extended \emph{a priori}
knowledge to build a proper model for the background \emph{and}
potential sources. The quality of a likelihood analysis is strongly
influenced by the quality of the chosen models.

This paper shows a way to incorporate the additional structure
information of extended sources into an analysis without the need for
prior knowledge about the source. To achieve this, a reliable and
powerful technique for quantifying structures of gamma-ray counts maps
is needed. The Minkowski functionals provide this technique. They are
well-studied tools from integral geometry \citep{SchneiderWeil:2008,
  Santalo:1976} and are widely used for both studying structures in
statistical physics \citep{Mecke:1998, MeckeStoyan:2000,
  SchroederTurketal:2009, SchroederTurketal:2010} and for pattern
analysis \citep{Mecke:1996, Becker:2003, MantzJacobsMecke:2008}. They
were already successfully applied in astronomy
\citep[e.g.][]{MeckeBuchertWagner:1994}. They were used to investigate
point processes in cosmology and the large-scale structure of the
universe \citep{KerscherMeckeSchuecker:2001,
  KerscherMecke:2001,Colombi:2000} and as probes of non-Gaussianity in
the cosmic microwave background
\citep{Schmalzing:1999,Gay:2012,Ducout:2013}.

This work uses Minkowski functionals for the first time to detect
sources in gamma-ray astronomy by quantifying structures in gamma-ray
counts maps. The paper is organized as follows:
section~\ref{StructureCharacterization} introduces the Minkowski
functionals in detail and describes the quantification of the
structure of the counts map. The morphometric analysis, i.e., the
actual source detection via structure quantification, is presented in
section~\ref{SourceDetection}. The way to express the background
morphology and define a compatibility of the structure of a measured
counts map with the background is explained in
section~\ref{GlobalNullHypothesisTest} using a global null
hypothesis. The technique is extended in
section~\ref{LocalMinkowskiSkyMaps} to local structure deviations
found with Minkowski sky maps, which allows one to resolve and
localize the gamma-ray sources. Section~\ref{SimulatedData} applies
the analysis to simulated data. Finally, the results for counts maps
observed with H.E.S.S. are given in section~\ref{HESSData}.


\section{Structure characterization}
\label{StructureCharacterization}

This section describes the structure characterization of a gamma-ray
counts map. While similar methods may be used to quantify the
morphology of extended gamma-ray sources, this is not the subject of
this paper.
Although the following structure analysis has not yet been applied in
gamma-ray astronomy, it is often used in integral geometry and in
statistical physics \citep{SchneiderWeil:2008, Mecke:1998,
  MeckeStoyan:2000}.

\begin{figure}[tb]
  \centering
  \subfigure[][]{%
    \includegraphics[width=0.48\linewidth]{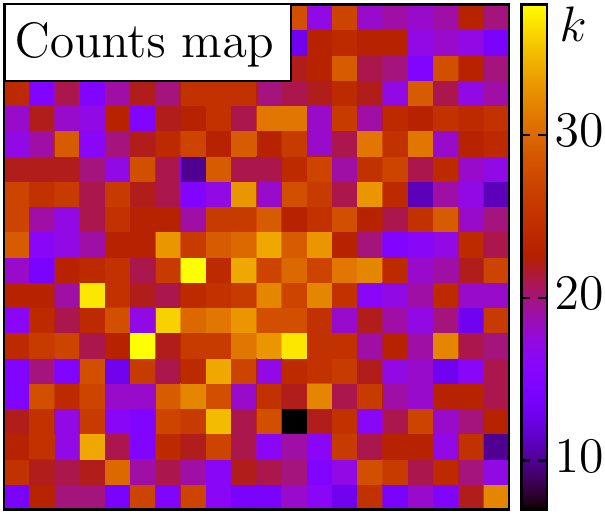}%
    \label{Minkowski_functionals_count_map}%
  }%
  \hspace{8pt}%
  \subfigure[][]{%
    \includegraphics[width=0.48\linewidth]{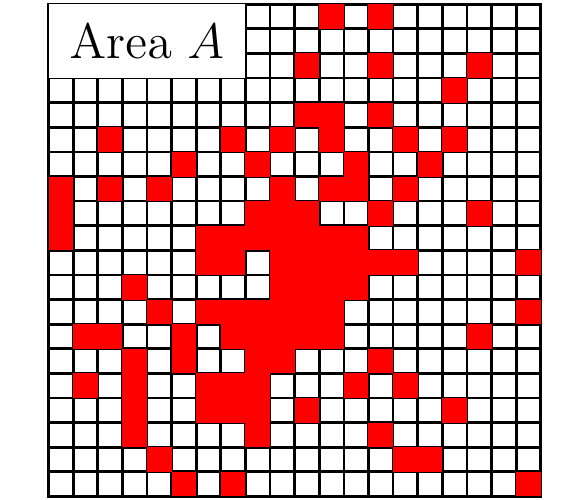}%
    \label{Minkowski_functionals_A}%
  }\\
  \subfigure[][]{%
    \includegraphics[width=0.48\linewidth]{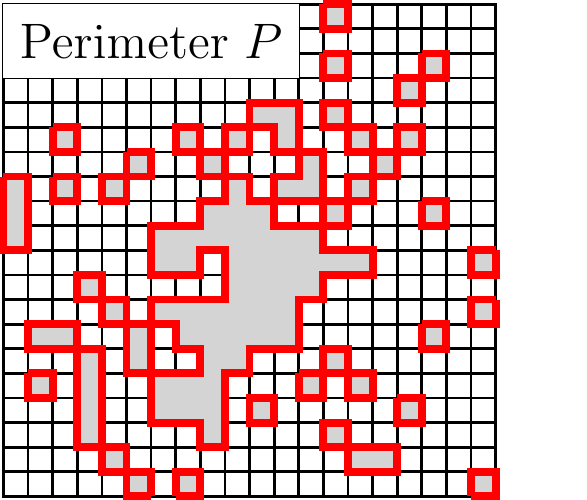}%
    \label{Minkowski_functionals_P}%
  }%
  \hspace{8pt}%
  \subfigure[][]{%
    \includegraphics[width=0.48\linewidth]{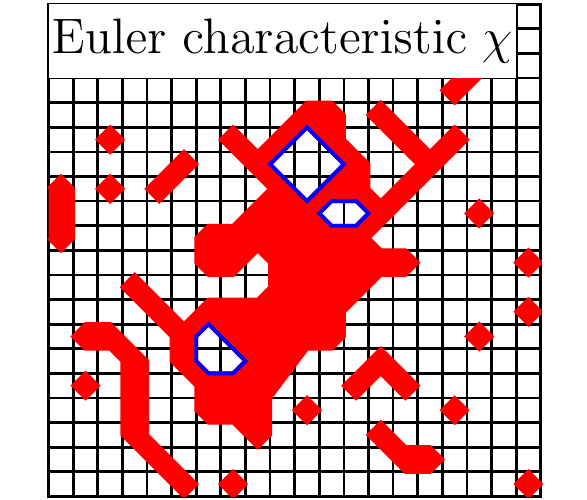}%
    \label{Minkowski_functionals_C}%
  }
  \caption{Structure quantification via Minkowski functionals. (a)
    Counts map, simulated Poisson-distributed random number of counts
    $k$. To characterize the morphology, the image is turned into a
    black-and-white image via thresholding --- see (b); the three
    Minkowski functionals are then evaluated for the b/w image. (b)
    Area $A$. (c) Perimeter $P$. (d) Euler characteristic $\chi$.}
  \label{Minkowski_functionals}
\end{figure}

A gray-scale image, here the counts map, is turned into a
black-and-white (b/w) image~\citep{Mecke:1996}. For each threshold
value $\rho$, all pixels with counts $k \geq \rho$ are set to black,
the others remain white --- see Fig.~\ref{Minkowski_functionals}. The
structure of the image is then analyzed as a function of the threshold
$\rho$.

The structure of each b/w image is quantified by the Minkowski
functionals\footnote{Other names are valuations, quermaßintegrals,
  intrinsic volumes, or Hadwiger measures.}. In two dimensions there
are three of them. They are proportional to well-known geometric
quantities: the area~$A$ of the black pixels, their perimeter~$P$, and
the Euler characteristic~$\chi$, which is the integral of the Gaussian
curvature. It is a topological constant; for closed domains it is
given by the number of components minus the number of
holes. Figure~\ref{Minkowski_functionals} visualizes how a counts
map~\subref{Minkowski_functionals_count_map} is turned into a b/w
image~\subref{Minkowski_functionals_A}, which is then quantified by
Minkowski
functionals~\subref{Minkowski_functionals_A}-\subref{Minkowski_functionals_C}.
The area as a function of the threshold contains the knowledge about
the number of counts. However, it does not supply any information
about their arrangement, for which additional information is provided
by the perimeter and the Euler characteristic.

The Minkowski functionals are powerful shape measures. Because of
their additivity and continuity, they are robust against noise and
have short computation times. There are several linear time algorithms
for calculating the area, perimeter, and Euler characteristic
\citep[e.g.][]{MantzJacobsMecke:2008,SchroederTurketal:2009} and for
$\unit[3]{D}$ data
\citep[e.g.][]{Arns:2010,SchroederTurketal:2010,IEEE:2012}.
\begin{table}[tb]
  \caption{Look-up table for Minkowski functionals: the functional
    values of area $A$, perimeter $P$, and Euler characteristic $\chi$
    are assigned to each 2x2~neighborhood of the image. The unit of
    length is the edge-length of a pixel. Similar data can be found
    in \citet{Mecke:1996} and \citet{MantzJacobsMecke:2008}.}
  \label{MinkowskiFunctionalsLookUpTable}
  \centering
  \begin{tabular}{r c     c c c     r c     c c c}
    \toprule
    Conf. &  & $A$ & $P$ & $\chi$ &  Conf.&  & $A$ & $P$ & $\chi$ \\
    \cmidrule(rl){1-5}\cmidrule(rl){6-10}
    1  & \includegraphics[height=9pt]{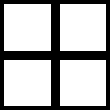}  & 0   & 0 & 0    & 9  & \includegraphics[height=9pt]{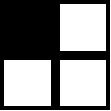}   & 1/4 & 1 & 1/4   \\    
    2  & \includegraphics[height=9pt]{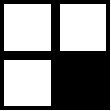}  & 1/4 & 1 & 1/4  & 10 & \includegraphics[height=9pt]{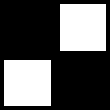}  & 1/2 & 2 & -1/2  \\
    3  & \includegraphics[height=9pt]{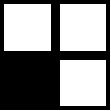}  & 1/4 & 1 & 1/4  & 11 & \includegraphics[height=9pt]{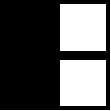}  & 1/2 & 1 & 0     \\
    4  & \includegraphics[height=9pt]{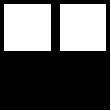}  & 1/2 & 1 & 0    & 12 & \includegraphics[height=9pt]{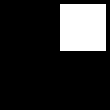}  & 3/4 & 1 & -1/4  \\
    5  & \includegraphics[height=9pt]{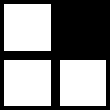}  & 1/4 & 1 & 1/4  & 13 & \includegraphics[height=9pt]{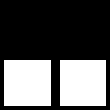}  & 1/2 & 1 & 0     \\
    6  & \includegraphics[height=9pt]{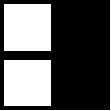}  & 1/2 & 1 & 0    & 14 & \includegraphics[height=9pt]{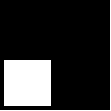}  & 3/4 & 1 & -1/4  \\
    7  & \includegraphics[height=9pt]{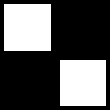}  & 1/2 & 2 & -1/2 & 15 & \includegraphics[height=9pt]{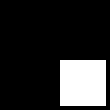}  & 3/4 & 1 & -1/4  \\
    8  & \includegraphics[height=9pt]{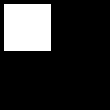}  & 3/4 & 1 & -1/4 & 16 & \includegraphics[height=9pt]{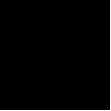}  & 1   & 0 & 0     \\
    \bottomrule
  \end{tabular}
\end{table}
The straightforward algorithm used here is based on
Table~\ref{MinkowskiFunctionalsLookUpTable}. The image is decomposed
into $2\times2$~neighborhoods. The value of the Minkowski functionals
is assigned to each of the 16~possible configurations. Because of
their additivity, the sum of the local contributions yields their
global value. The unit of length is defined as the edge-length of a
single pixel, thus the area of a pixel is one. To avoid multiple
countings when iterating over the whole image, only that part may
contribute which is unique to a $2\times2$~neighborhood, i.e., each
quarter of the four pixels next to the center. For example, a single
black pixel has area and Euler characteristic one and perimeter
four. However, when iterating over the image, it will appear in four
different 2x2~neighborhoods, namely configurations two, three, five,
and nine. Thus, Table~\ref{MinkowskiFunctionalsLookUpTable} assigns to
each of them area and Euler characteristic one fourth and perimeter
one. The white pixel in configuration 15 can be interpreted as part of
a hole; it contributes negatively to the Euler characteristic. In
configurations seven and ten in
Table~\ref{MinkowskiFunctionalsLookUpTable} the black pixels sharing
only a vertex are chosen to be connected. If they were disconnected,
the weights for the Euler characteristic would be positive. The choice
is arbitrary, as long as the probability distribution for the Euler
characteristic is calculated consistently. However, connecting them
helps to distinguish a single cluster of black pixels from two domains
distant from each other.\footnote{If a marching square algorithm is
  used to find a more complex triangulation of the domain of black
  pixels, the weights for area and perimeter have to be adjusted --
  see \citet{MantzJacobsMecke:2008}. The probability distributions for
  the Minkowski functionals have to be calculated
  consistently. However, no significant effect on the final results
  has yet been observed.}

The choice of boundary conditions has a strong influence on the
structure quantification and its efficiency~\citep{Stoyan:1987}.
Throughout this work, closed boundary conditions are applied. This
means all pixels outside the window of observation are set to white,
thus all domains are closed. A discussion of the different impacts and
drawbacks of various boundary conditions is beyond the scope of this
paper and will follow in future publications.

Hadwiger's completeness theorem ensures that the Minkowski functionals
provide a robust and comprehensive morphology analysis, i.e., the
Minkowski functionals form a complete basis of all valuations that are
defined on unions of convex sets and are motion invariant, additive,
and at least continuous on convex sets~\citep{Hadwiger:1957}. To
quantify anisotropy, they can be generalized to tensor
valuations~\citep[e.g.][]{SchneiderWeil:2008, SchroederTurketal:2009,
  SchroederTurketal:2010}.


\section{Source detection}
\label{SourceDetection}

\subsection{Global null hypothesis test}
\label{GlobalNullHypothesisTest}

Structural deviations from the background morphology are to be
detected. Therefore, the characteristic structure of a background
measurement has to be known.
Because the Minkowski functionals quantify the morphology, the
probability distribution~$\mathcal{P}$ of their functional value~$X
\in \{A,P,\chi\}$ for a counts map of a background measurement is
needed. The first step is to choose a suitable background model.

In general, determining~$\mathcal{P}(X)$ is the most complex task in
the morphometric analysis. A reasonable model for background counts in
VHE gamma-ray astronomy is the assumption of homogeneously and
isotropically Poisson-distributed counts in each bin of a sky map with
equal area bins. This is because most background events in
ground-based VHE gamma-ray astronomy are caused by VHE hadrons. These
hadrons loose their direction correlations in interstellar magnetic
fields and arrive at the Earth as a uniform flux of VHE particles from
every direction. For a real measurement the homogeneous and isotropic
background will of course be distorted by detector effects and
non-uniform exposure of the sky. As we show in
section~\ref{DetectorAcceptanceCorrection}, the data can be corrected
for these effects, i.e., the typical structure of a background
measurement can be derived from the structure of a pure homogeneous
and isotropic Poisson background.\footnote{The choice of bin size is
  free, because a uniform Poisson field remains a homogeneous Poisson
  field for any chosen bin size and an arbitrary point spread
  function; the null hypothesis will be unchanged. However, to
  quantify the actual source morphology and not random noise, the bin
  size should be adjusted to the point spread function.}

\begin{figure}[tb]
  \centering
  \subfigure[][]{%
    \includegraphics[width=0.49\linewidth]{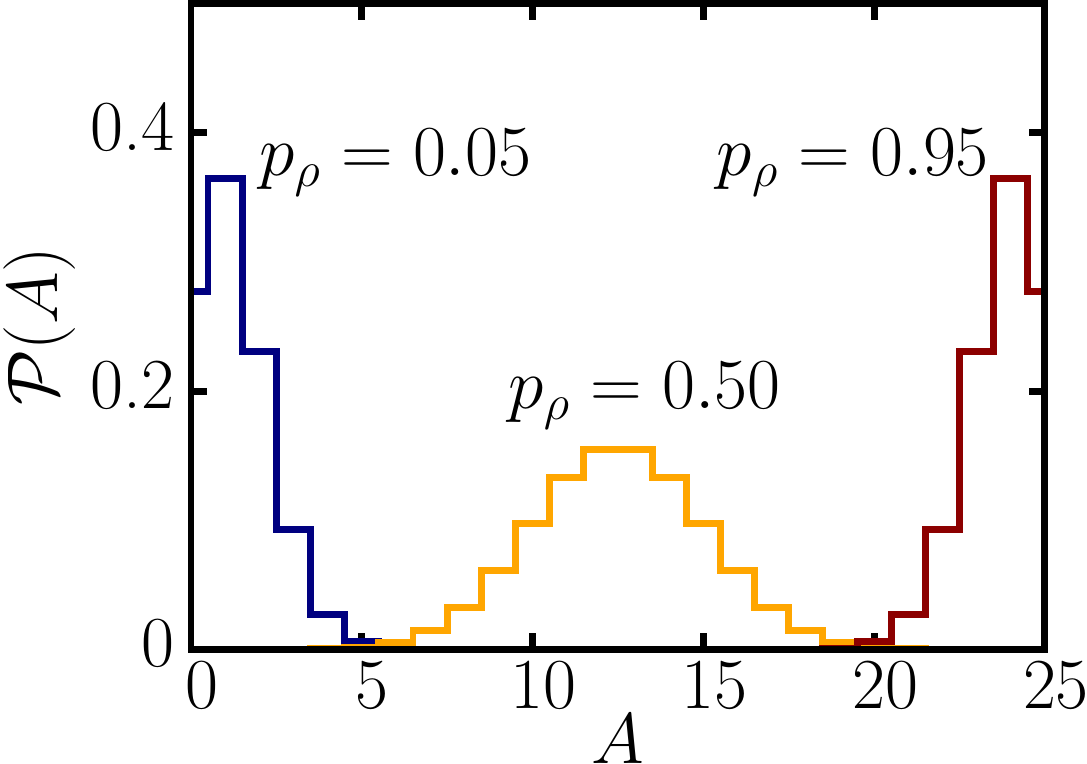}%
    \label{structure_distributions_P_A}%
  }%
  \hspace{4pt}%
  \subfigure[][]{%
    \includegraphics[width=0.49\linewidth]{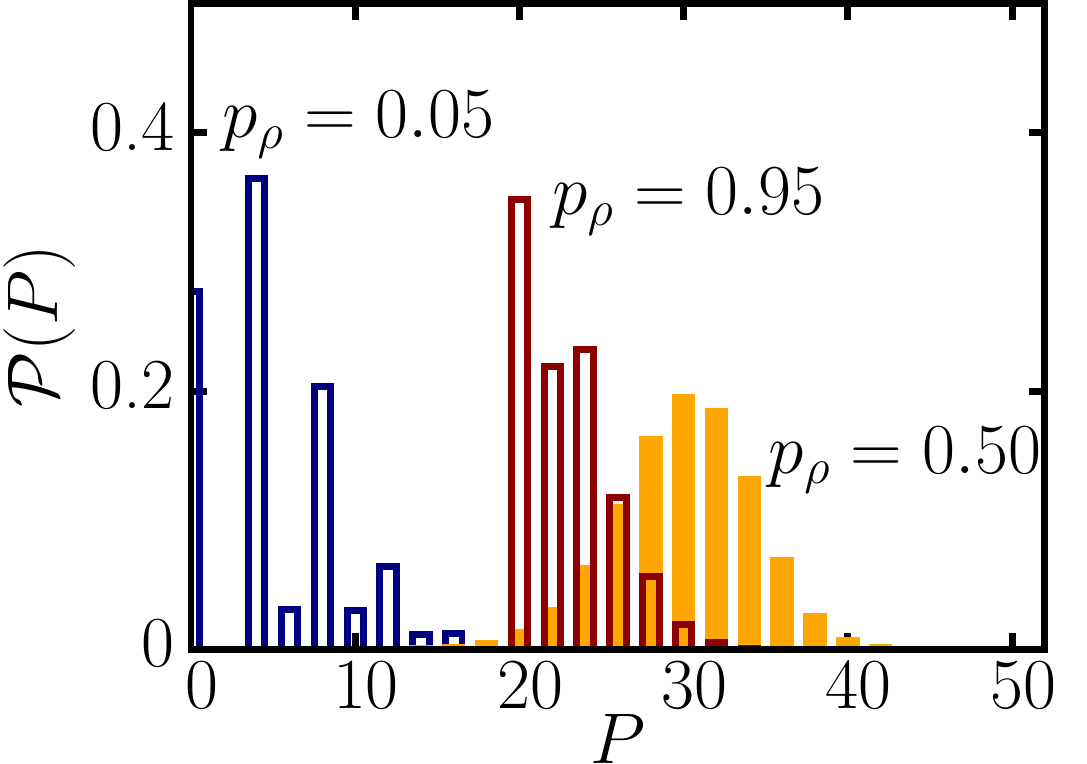}%
    \label{structure_distributions_P_P}%
  }\\
  \subfigure[][]{%
    \includegraphics[width=0.49\linewidth]{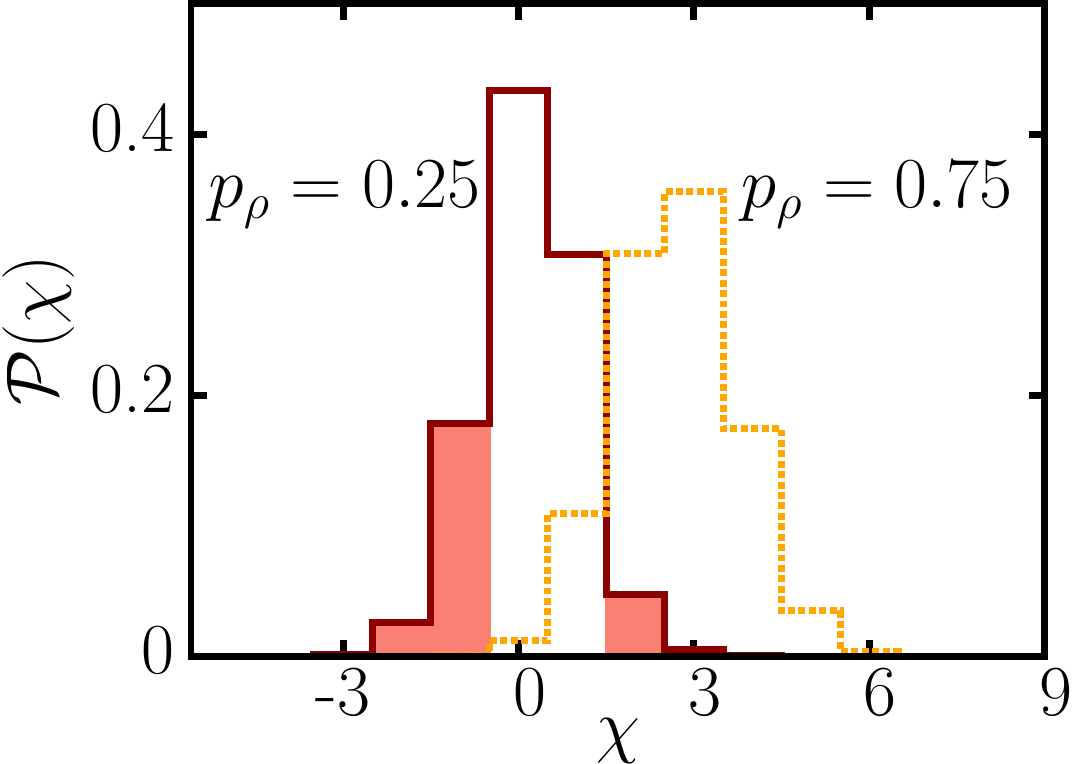}%
    \label{structure_distributions_P_C}%
  }%
  \hspace{4pt}%
  \subfigure[][]{%
    \includegraphics[width=0.49\linewidth]{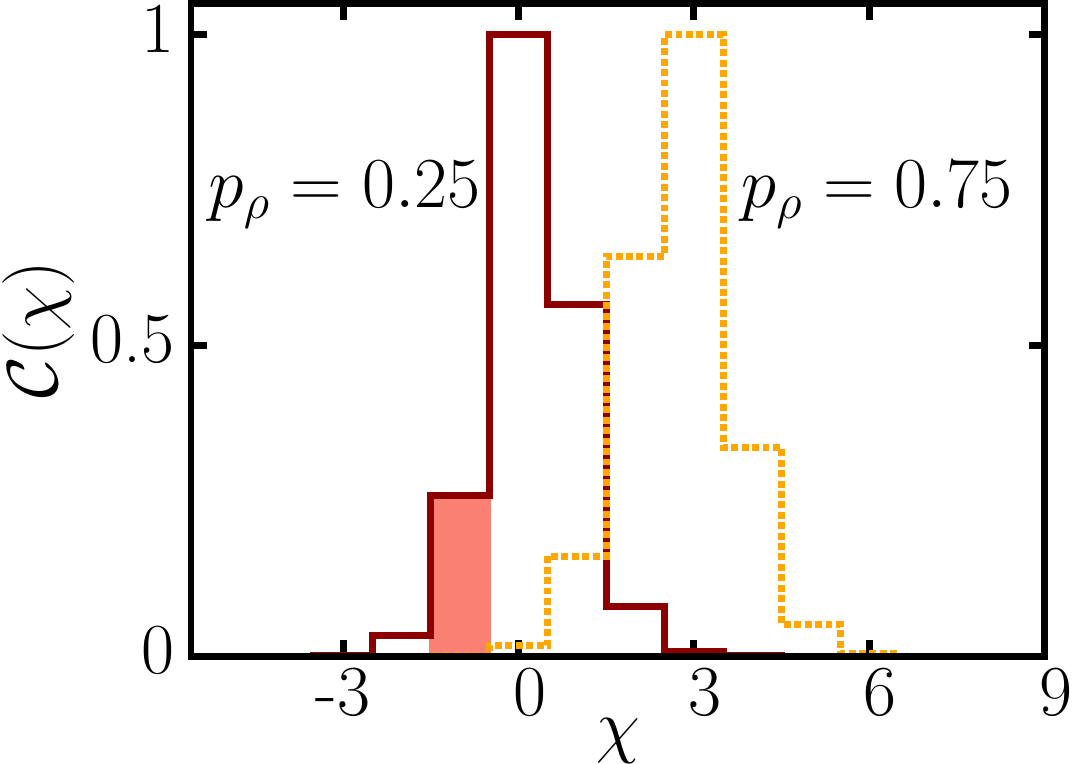}%
    \label{structure_distributions_C_C}%
  }
  \caption{Structure of background noise for a $5\times5$ bin field at
    different thresholds $\rho$ resulting in different probabilities
    $p_{\rho}= \sum_{i=\rho}^{\infty}{\lambda^i\, e^{-\lambda}/i!}$ for a black pixel. Probability distributions
    $\mathcal{P}$ of (a) area $A$; (b) perimeter $P$; (c) Euler
    characteristic $\chi$. (d) Compatibility $\mathcal{C}$ of Euler
    characteristic $\chi$, with $\mathcal{C}(\chi=-1)$ the sum of all
    probabilities $\mathcal{P} \leq \mathcal{P}(\chi=-1)$, which is
    visualized by the equally large colored regions beneath the graphs
    in (c) and (d)~-~mind the different scales.}
  \label{structure_distributions}
\end{figure}

The likelihood for the area~$A$ of the black pixels is a suitable
introductory example; the probability distribution for each
threshold~$\rho$ is given by the binomial distribution
\begin{align*} 
  \mathcal{P}(A) &= \mathrm{Binom}(A,p_{\rho},N^2) \, ,
\end{align*}
where~$A$ is the number of black pixels, $N^2$ the number of pixels in
an~$N \times N$ sky map, and~$p_{\rho}$ the probability that a pixel
is black, i.e., there are more counts or their number is equal to the
threshold~$\rho$. Assuming a Poisson background with an expected
number of counts per bin of~$\lambda$, $p_{\rho}$ is given by
\begin{align}
  p_{\rho} &= \sum_{i=\rho}^{\infty}{\frac{\lambda^i}{i!} e^{-\lambda}} \, .
  \label{eq:p.rho}
\end{align}
The probability distributions $\mathcal{P}(A)$, $\mathcal{P}(P)$, and
$\mathcal{P}(\chi)$ of the area, perimeter, and Euler characteristic,
respectively, are plotted in Fig.~\ref{structure_distributions}~(a-c)
as a function of the probability~$p_{\rho}$ that a pixel is above the
threshold~$\rho$. For small $N$ these distributions can be found by
evaluating all possible b/w images and inferring the distribution by
counting equivalent pixel configurations.

Gamma-ray sources can be detected by looking for structures, which are
very unlikely to be found if the hypothesis of a pure background
measurement was true, i.e., assuming that there are only background
events within the observation window. A probability measure that a
given counts map with a structural value~$X$ is compatible with this
hypothesis may be defined as follows: The compatibility
\begin{align}
  \mathcal{C}(X) &= \sum_{\mathcal{P}(X_i) \le \mathcal{P}(X)} \mathcal{P}(X_i)
\end{align}
is the probability that a less likely structure
appears. Figure~\ref{structure_distributions} (d) shows the
compatibility~$C(\chi)$ for the Euler characteristic from
Fig.~\ref{structure_distributions} (c).

The compatibility is defined following the scheme given in
\citet{effHypoTest} to construct a most efficient hypothesis
test. The form used here may be derived from the general scheme given
in the paper by setting the supremum of alternative hypotheses to $1$,
i.e., imposing no constraints on alternative hypotheses.
The hypothesis of a pure background measurement is rejected if the
compatibility is lower than $0.6\cdot10^{-6}$. This hypothesis
criterion is adjusted to the commonly used $\unit[5]{\sigma}$
deviation; a normally distributed random variable deviates from the
expected value by at least $\unit[5]{\sigma}$ with a probability of
approximately $0.6\cdot10^{-6}$.

Instead of dealing with tiny compatibilities, it is often more
convenient to use the logarithm of this likelihood value or to define
the deviation strength
\begin{align}
  \mathcal{D} :=& - \log_{10}(\mathcal{C}) \, .
\end{align}
The conversion between compatibility~$\mathcal{C}$ and
standard deviation~$\sigma$ is given by $\mathcal{C}(\sigma) = \mathrm{erfc}
\left(\sigma/\sqrt{2}\right)$,
where $\mathrm{erfc}(x) = \frac{2}{\sqrt{\pi}} \int_x^\infty{\exp(-t^2) \;
  \mathrm{d}t}$ is the error function.

\begin{figure}[tb]
  \centering
  \includegraphics[width=\linewidth]{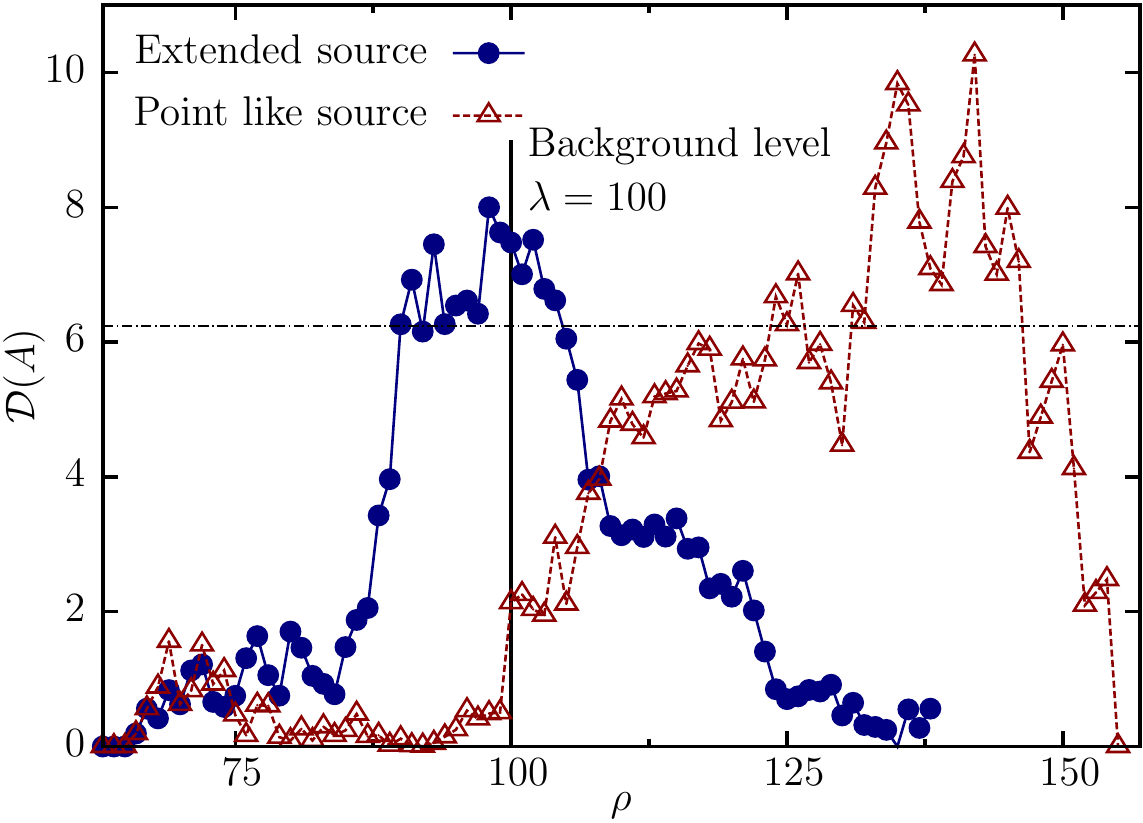}
  \caption{Deviation strength $\mathcal{D}(A)$ for structure
    quantification via area~$A$ as a function of the threshold~$\rho$
    for an extended and for a point source of equal flux within an
    observation window of $100 \times 100$ bins.
  }
  \label{threshold_dependence_example}
\end{figure}

Figure~\ref{threshold_dependence_example} depicts a structure analysis
of both a rather extended and of a more point-like source,
investigating simulated data of Poisson-distributed random number of
counts. The deviation strength~$\mathcal{D}$ is plotted over the
threshold~$\rho$. Because there are thresholds for both sources for
which the deviation strength is greater than $6.2$, the null
hypothesis of a pure background measurement can be rejected in either
case. The results differ for the two sources, but their total flux is
equal, and reveals basic information about the extension of the
sources. The point source cannot be detected at thresholds near the
background level~$\lambda$ and is only apparent for high thresholds,
i.e., only bins with high counts contribute to the detection. For the
extended source only counts in the order of the magnitude of the
background fluctuations are present in the sky map. However, as the
number of bins with slightly increased counts exceeds the typical
background predictions, the hypothesis is nevertheless rejected.
 
\subsection{Local Minkowski sky maps}
\label{LocalMinkowskiSkyMaps}

So far, it is possible to scan the entire field of view (FoV) of an
observation for additional structures w.r.t. the expected
background. To localize the deviations and thus locate gamma-ray
sources and gain insight into their extension and morphology, more
information is needed. Owing to their motion invariance, the scalar
Minkowski functionals cannot be used directly to localize
structures.\footnote{While the generalization to motion covariant
  tensorial valuations might provide this insight, these
  generalization will not be discussed in this paper. Instead a
  straightforward and easy approach to the problem is discussed here.}

Instead of analyzing the entire FoV with the methods introduced so
far, a small sliding window may be used. This window can be moved
across the FoV to study the local structure of the sky map in the
sliding window. With this approach one can construct Minkowski sky
maps from the counts maps that are just as useful as significance maps
constructed using the approach from \citet{lima}, but which provide
the additional sensitivity from the structure information used to
determine the deviation strength.

Since the maximum deviation strength is assigned to a pixel, a trial
factor must be added to the result to avoid overestimating the
significance of the found deviations~\citep{Goering:2012}. Each of the
different b/w images for different thresholds $\rho$ contributes a
trial to the search for structure deviations, and as the number of
trials increases, the probability to find a significant random
fluctuation increases as well. Assuming one looked for measurements
with a compatibility lower than $\alpha$, the probability to find such
a deviation in one image is $\alpha$. The probability to find no such
deviation in $t$ independent images is $(1-\alpha)^t$ and the
probability to find at least one such deviation in the image set is
$1-(1-\alpha)^t$.  Thus, the significance after $t$ independent trials
is linked to the pre-trial significance $\alpha$ via
\begin{equation}
  \alpha_t = 1-(1-\alpha)^t = t \alpha + \mathcal{O}(\alpha^2) \, .
  \label{eq:postTrial}
\end{equation}
Accordingly, for $\alpha \ll 1$ the influence of $t$ independent
trials may be approximated by the so-called trial factor, i.e., by
multiplying $\alpha$ with $t$. For the corresponding deviation
strength $\mathcal{D}_t$ this results in an offset of $\log_{10}(t)$
compared with the pre-trial deviation strength $\mathcal{D}$,
i.e., $\forall \mathcal{D} \gg 0: \mathcal{D}_t \approx \mathcal{D} -
\log_{10}(t)$.

Obviously, the different b/w images resulting from thresholding are
not statistically independent, because they originate from the same
intensity profile. If the structure of a b/w image at a certain
threshold can be interpolated from the structures of the b/w images of
the next higher and lower thresholds, it does not contribute a
separate trial to the search for deviations from the null
hypothesis. Therefore, if $t$ is set to the number of threshold steps,
Eq.~\eqref{eq:postTrial} will yield a conservative estimate of the
post-trial significance.

A rough estimate of the systematic error on the deviation strength
$\mathcal{D}$ introduced by ignoring the influence of trial factors is
given by $2 \log_{10}(N)$. This estimate is based on the fact that the
number of different b/w images after thresholding is equal to the
number of different gray levels in the original gray-scale image that
represents the gamma-ray counts map. There are at most $N^2$ different
gray levels in an image of $N \times N$ pixels and therefore at most
$N^2$ different trials may contribute to the search for structure
deviations; from $t \le N^2$ follows
\begin{equation}
  \mathcal{D}_t \approx \mathcal{D} -\log_{10}(t) \ge \mathcal{D} - 2\log_{10}(N) \, .
\end{equation}
Therefore, $\mathcal{D} - 2\log_{10}(N)$ may be used to compute a
conservative estimate of the post-trial deviation strength; for the
preceding discussion see \citet{Goering:2012}.

\begin{figure}[tb]
  \centering
  \includegraphics[width=\linewidth]{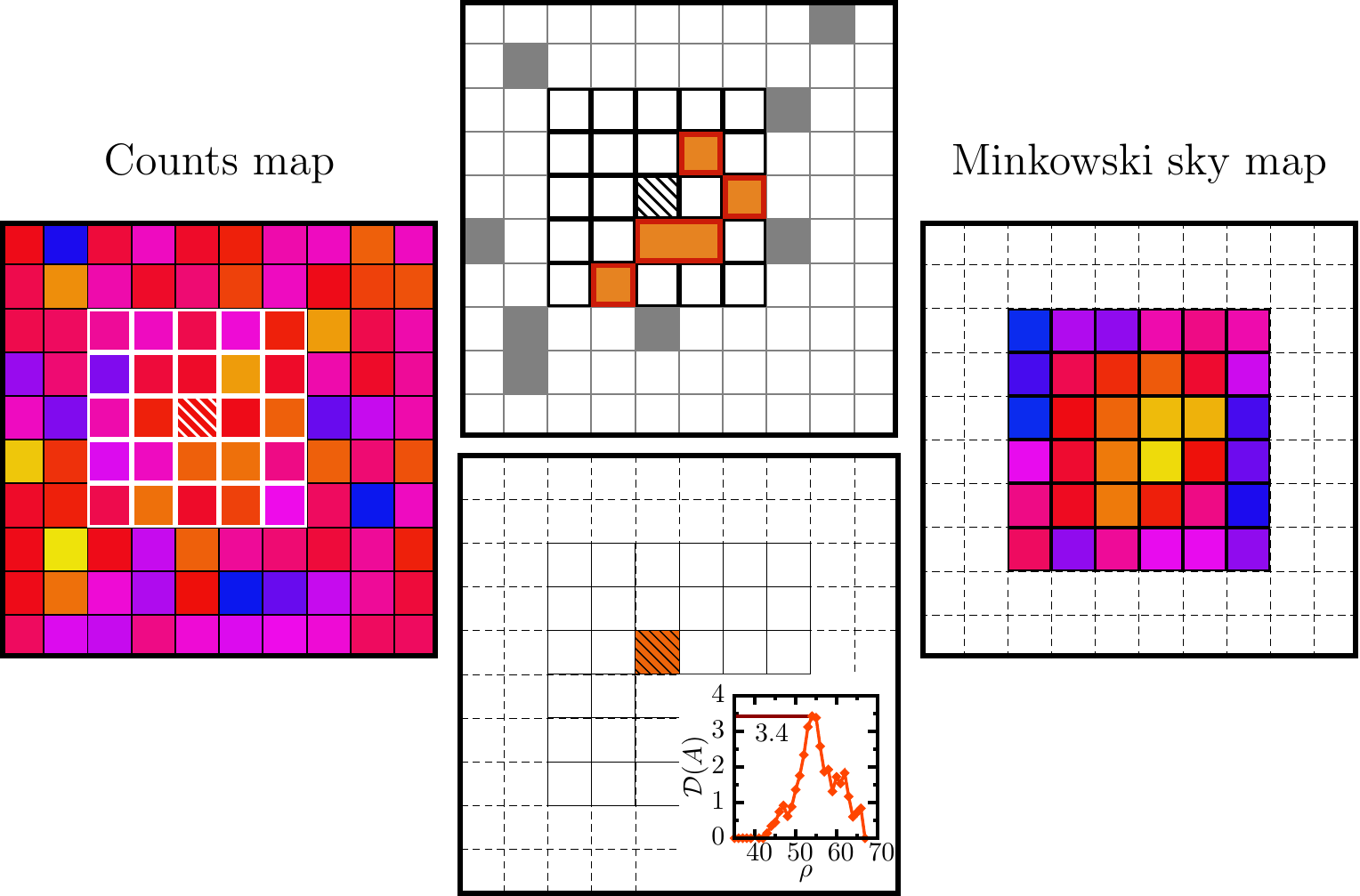}
  \caption{Computation of a Minkowski sky map. LEFT: a given counts
    map. MIDDLE: for each pixel the local structure within the
    sliding window is characterized (top) and the maximum deviation
    strength for all thresholds is assigned to it in the Minkowski sky
    map (bottom). RIGHT: iterating through all pixels provides the
    Minkowski sky map.}
  \label{computation_Minkowski_sky_map}
\end{figure}

Figure~\ref{computation_Minkowski_sky_map} depicts the concept of a
\textit{Minkowski sky map}. For each of the inner pixels of the counts
map the local structure is to be characterized (depicted on the
left). Therefore, the Minkowski functionals of a sliding window with
$N \times N$ pixels are evaluated (illustrated in the top picture in
the middle of Fig.~\ref{computation_Minkowski_sky_map}). The maximum
deviation strength for all thresholds~$\rho$ is assigned to the pixel
at the center of the sliding window (plotted in the bottom picture in
the middle). Iterating over all inner pixels for which the sliding
window is completely within the counts map provides the Minkowski sky
map.

The information content and intuitive interpretation of such Minkowski
sky maps can be enhanced by adding a sign to the deviation strength of
the different pixels. By choosing the sign to be negative if $A < N^2
p$, i.e., if there are fewer black pixels than expected, and positive
otherwise, the sign of a pixel shows if the local deviation is caused
by an overestimation of the background or by additional flux from
potential gamma-ray sources.  Minkowski sky maps detect local
structural deviations and depict them in an illustrative and
quantitative image, depicting the lack of trust in the hypothesis that
there are only background fluctuations.

Although the null hypothesis is tested locally, the background
assumption is a global null hypothesis, i.e., the expected number of
counts per bin $\lambda$ is chosen globally. If $\lambda$ is chosen
locally for every sliding window, structures larger than the sliding
window may result in an increased $\lambda$ and in no deviation at all
if the local structure is homogeneous and isotropic within the sliding
window.


\subsection{Simulated data}
\label{SimulatedData}

{ \subfiglabelskip=0pt
  
  \begin{figure}[tb]%
    \centering
    \subfigure[][]{%
      \includegraphics[width=0.48\linewidth]{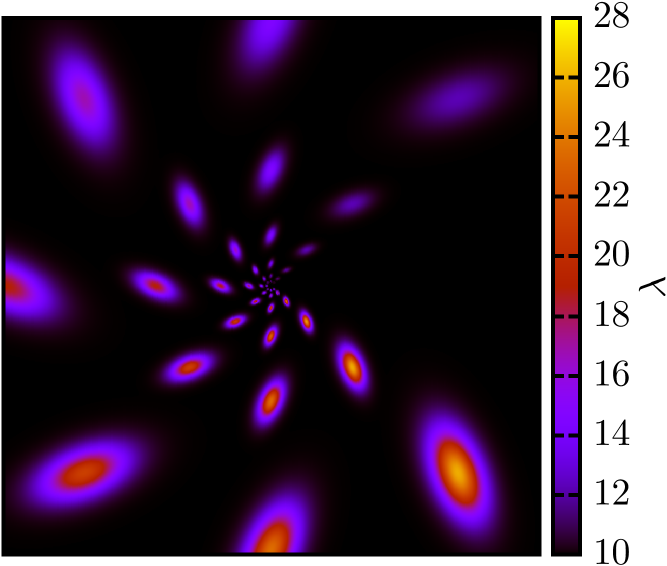}%
      \label{MinkowskiSkyMap_pix_intensity}%
    }%
    \hspace{8pt}%
    \subfigure[][]{%
      \includegraphics[width=0.48\linewidth]{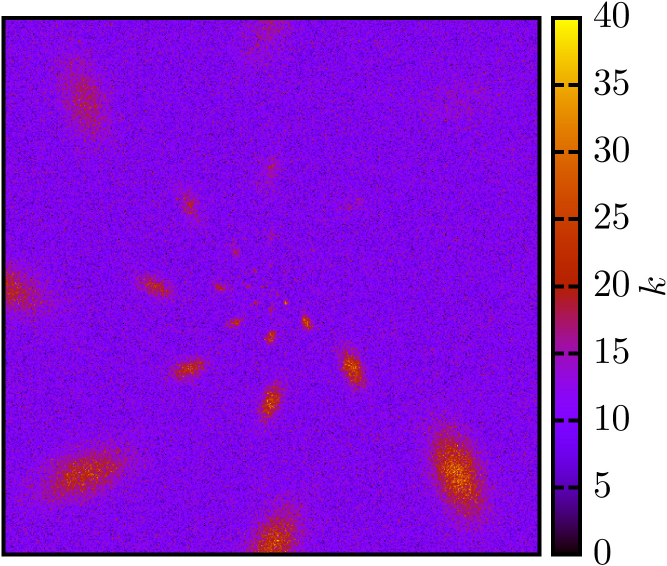}%
      \label{MinkowskiSkyMap_pix_count_map}%
    }\\
    \subfigure[][]{%
      \includegraphics[width=0.48\linewidth]{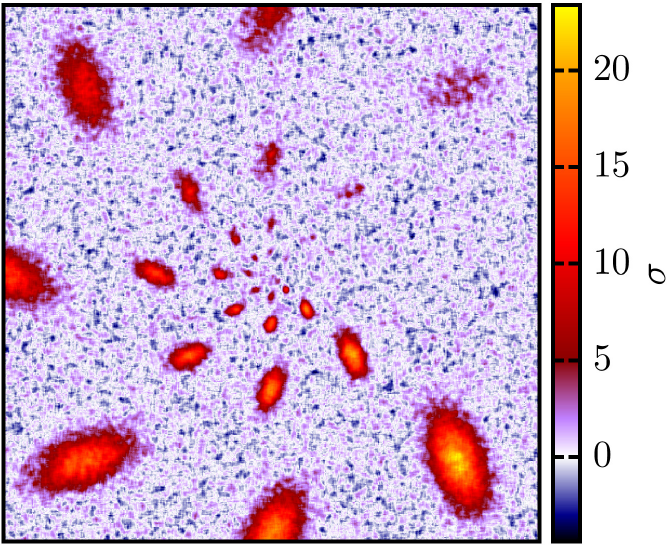}%
      \label{MinkowskiSkyMap_pix_significance}%
    }%
    \hspace{8pt}%
    \subfigure[][]{%
      \includegraphics[width=0.48\linewidth]{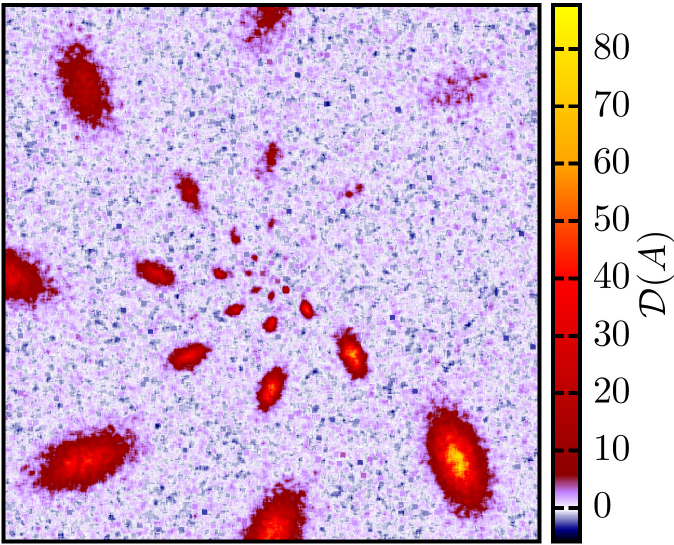}%
      \label{MinkowskiSkyMap_pix_MinkowskiSkyMap_A}%
    }\\
    \subfigure[][]{%
      \includegraphics[width=0.48\linewidth]{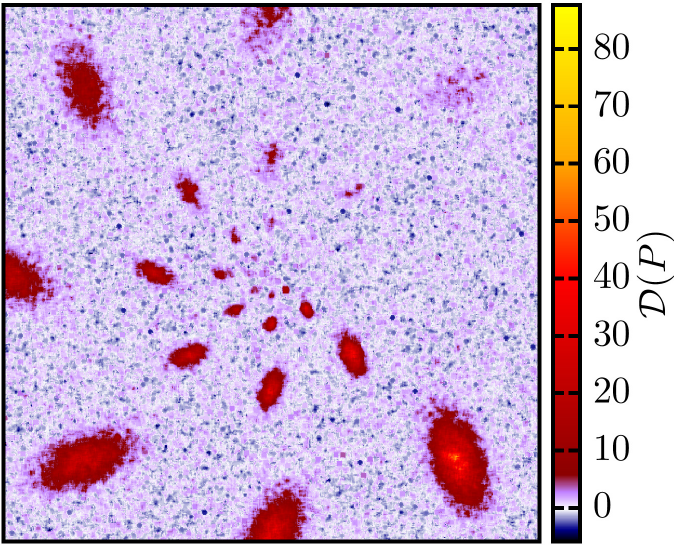}%
      \label{MinkowskiSkyMap_pix_MinkowskiSkyMap_P}%
    }%
    \hspace{8pt}%
    \subfigure[][]{%
      \includegraphics[width=0.48\linewidth]{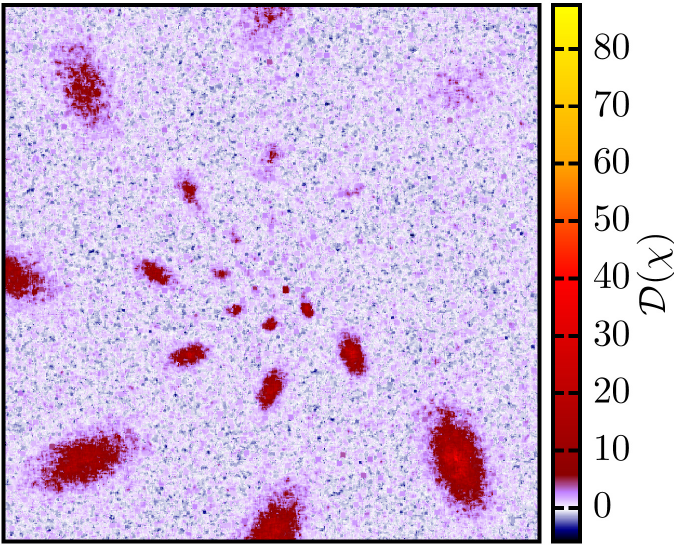}%
      \label{MinkowskiSkyMap_pix_MinkowskiSkyMap_C}%
    }
    \caption{Structure analysis of simulated sky maps.
      \subref{MinkowskiSkyMap_pix_intensity} Given intensity
      profile. \subref{MinkowskiSkyMap_pix_count_map} Simulated counts
      map. \subref{MinkowskiSkyMap_pix_significance} Significance map,
      evaluated with standard techniques, weighting the count
      excess. \subref{MinkowskiSkyMap_pix_MinkowskiSkyMap_A} Minkowski
      sky map; the structure is characterized with the area~$A$,
      \subref{MinkowskiSkyMap_pix_MinkowskiSkyMap_P} the
      perimeter~$P$, and
      \subref{MinkowskiSkyMap_pix_MinkowskiSkyMap_C} the Euler
      characteristic~$\chi$.}%
    \label{MinkowskiSkyMap_pix}%
  \end{figure}
}

It is now possible to detect, localize, and study the shape of
gamma-ray sources using the morphometric analysis with Minkowski sky
maps, introduced in the previous
section. Figure~\ref{MinkowskiSkyMap_pix} depicts an analysis of
exemplary simulated data.
Figure~(a) shows the given intensity profile of the test pattern
sources, the expected number of counts per bin caused by the chosen
sources. The test pattern consists of Gaussian-shaped extended sources
with a ratio of the semi-major axes of $2$. The peak intensities of
the sources along a central ray from the image center are equal. The
sizes of sources with the same distance from the center are equal and
the peak intensities are increasing in counter-clockwise
direction. Plot~(b) is the simulated counts map analyzed in~(c-f). In
figure~(c) the standard analysis technique of \citet{lima} as used by
the H.E.S.S. experiment \citep[cf.][]{crab} is applied to the counts
map~(b) for comparison. Figures~(d-f) are the Minkowski sky maps with
a sliding window size $5\times5$; the structure is quantified by
either the area~$A$, the perimeter~$P$, or the Euler
characteristic~$\chi$. The same $5 \times 5$ window is used as
\emph{on}-region for the significance determination using
\citet{lima}.

The first apparent result drawn from Fig.~\ref{MinkowskiSkyMap_pix} is
that the single functionals and the standard analysis are similarly
sensitive to sources; area, perimeter, and Euler characteristic are
comparably competitive in finding gamma-ray signals.


\section{H.E.S.S. data}
\label{HESSData}

\subsection{Detector acceptance correction}
\label{DetectorAcceptanceCorrection}

An analysis of real data must always take into account the detector
acceptance. After modeling the camera acceptance, the effect must be
corrected for to regain an isotropic and homogeneous structure for
background measurements. For each bin~$i$ only a fraction $f_i$ of the
signals are expected to be detected. If each bin is simply weighted
with $1/f_i$, fractional photon counts will occur that destroy the
Poisson structure of the sky map.

The null hypothesis, that is, that for an ideal camera acceptance
there is only background noise with intensity~$\lambda$, allows an
acceptance correction, which preserves the Poisson structure. Because
only the fraction $f_i$ of the events in bin~$i$ are detected, the
actual background intensity~$\lambda_i = f_i \cdot \lambda$ varies
with each bin. Following the null hypothesis, that the number of
counts is a Poisson-distributed random variable with mean~$\lambda_i$,
the original random process with mean~$\lambda$ can be regained if a
new Poisson-distributed random variable with mean~$\lambda_i^+ =
(1-f_i) \cdot \lambda$ is added.

In the center of the field of view, where $f_i \approx 1$, the number
of counts remains effectively unchanged. For $f_i \ll 1$ the
additionally created pseudo-photon counts may cover features, but they
never introduce additional structural deviations from the homogeneous
isotropic Poisson field because they fulfill the given null hypothesis
by construction. Covering regions of low acceptance with a layer of
pseudo-events corresponds to the fact that the instrument is less
sensitive to signals in these regions than in regions with high
acceptance -- see also~\citet{Goering:2008}, \citet{Klatt:2010}, and
\citet{Goering:2012}.

\subsection{H.E.S.S. source}
\label{HESSsource}

With the methods discussed so far, we can analyze experimental data of
ground-based VHE gamma-ray telescopes such as the
H.E.S.S. experiment. Because the morphometric analysis is targeted at
extended structures, \mbox{\textnormal{RX J$1713.7\!-\!3946$}} was
chosen for an exemplary analysis. This source has one of the largest
angular diameters of the sources seen with H.E.S.S. and its morphology
has already been studied in detail and is well known \citep{rxj1713},
which makes it a suitable benchmark for more detailed analyses.

The data set used for analysis corresponds to the data used in
\citet{rxj1713}, but instead of the Hillas-based event reconstruction
discussed there, the advanced event reconstruction based on a
likelihood-model fit presented in \citet{modelPP} was used to compute
the list of reconstructed events from the recorded camera
images. These reconstructed events were used to fill the binned counts
map, which is the main input of the morphometric analysis.

The second input needed is an acceptance map describing the spatial
sensitivity of the given observations, which is used to perform the
acceptance correction discussed above. This map was created using
standard H.E.S.S. analysis tools. In particular, the so-called
$\unit[2]{D}$ acceptance model discussed in \citet{habilMathieu} was
used. The overall background level $\lambda$ was determined from the
counts map by excluding all regions containing known sources of VHE
gamma-rays and computing the mean of the counts in the remaining bins
normalized by the corresponding acceptance.

Figure~\ref{HESS_real_data} shows the resulting Minkowski sky maps of
a morphometric analysis based on the area $A$, the perimeter $P$, and
the Euler characteristic $\chi$. The underlying counts and acceptance
maps used for this analysis were created using square bins of
$\unit[0.02]{\degree}$ width. The resulting sky maps clearly show
\mbox{\textnormal{RX J$1713.7\!-\!3946$}} and agree well with the
results of the standard H.E.S.S. analysis given in
\citet{rxj1713}. This demonstrates that an analysis based on the
structure of the measured counts map is indeed possible and provides
results comparable with those of well-established tools.

{ \subfiglabelskip=0pt
  \begin{figure}[tb]%
    \centering
    \subfigure[][]{%
      \includegraphics[width=0.32\linewidth]{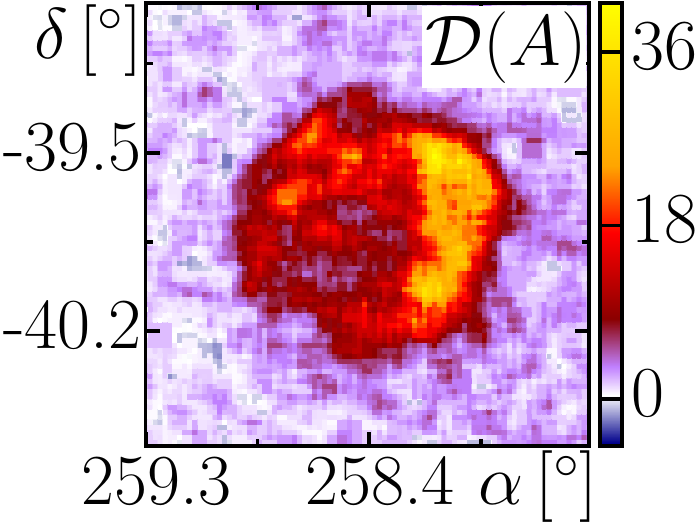}%
      \label{HESS_real_data_A}%
    }%
    \hspace{4pt}%
    \subfigure[][]{%
      \includegraphics[width=0.32\linewidth]{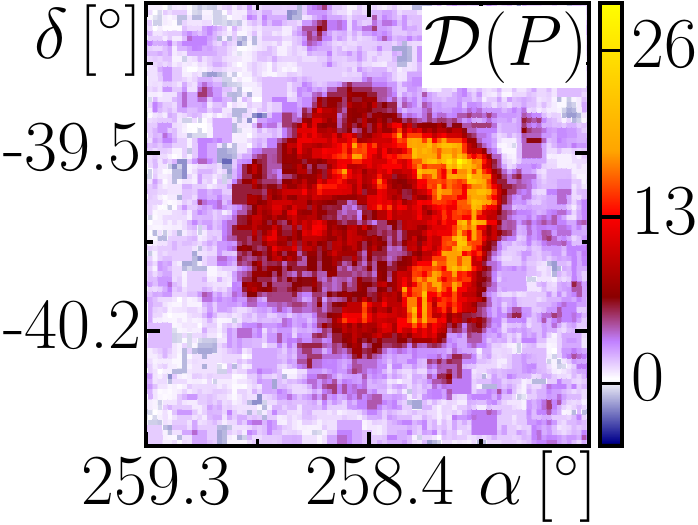}%
      \label{HESS_real_data_P}%
    }%
    \hspace{4pt}%
    \subfigure[][]{%
      \includegraphics[width=0.32\linewidth]{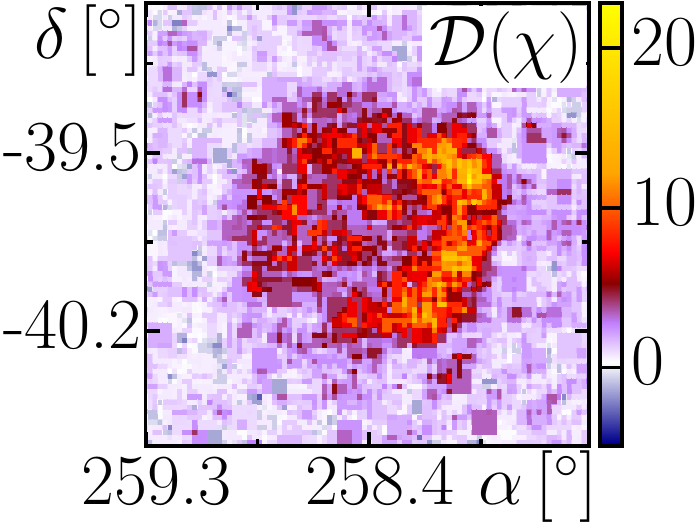}%
      \label{HESS_real_data_C}%
    }
    \caption{Minkowski sky maps of \mbox{\textnormal{RX
          J$1713.7\!-\!3946$}}. The morphometric analysis is based on
      (a)~area $A$, (b)~perimeter $P$, and (c)~Euler characteristic
      $\chi$.}%
    \label{HESS_real_data}%
  \end{figure} 
}


\section{Conclusion and outlook}
\label{ConclusionAndOutlook}

The introduced morphometric analysis provides a novel approach to data
analysis in VHE gamma-ray astronomy. In contrast to the commonly
applied hypothesis test by \citet{lima} to detect a significant excess
of gamma-ray counts on top of the expected background, it allows one
to incorporate additional morphometric information into the analysis
of gamma-ray counts maps. Still, the underlying model depends only on
$\lambda$, the expected background level of a given measurement, and
there is no need for a priori modeling of potential sources, as
opposed to an analysis based on a likelihood fit of a comprehensive
model to the data, as commonly used in satellite-based gamma-ray
experiments, for instance, Fermi/LAT \citep{fermilat}.
\citet{Klatt:2012} provided a short introduction to the shape analysis
of counts maps.

The morphometric analysis is based on the characterization of the
typical structure of a pure homogeneously and isotropically
Poisson-distributed background counts map. This typical structure is
determined using the Minkowski functionals --- morphometric valuations
from integral geometry. Significant deviations from the typical
background structure in measured gamma-ray counts maps can be used to
detect gamma-ray sources in the same way as significant excess counts
are used in analyses based on \citet{lima}.

With our basic ideas, it is possible today to qualitatively reproduce
analysis results based on well-established tool chains. To quantify
the agreement of the different analysis techniques and potential
sensitivity gains of the new morphometric approach, more in-depth
studies are required. Still, there are strong indications that the
presented methods will lead to a significant sensitivity gain in the
foreseeable future. Current proof-of-concept studies show an
impressive performance and a similar sensitivity of the three
Minkowski functionals to deviations from the background
structure. Combining the different functionals to refine the
background characterization may well lead to a significant sensitivity
boost. Furthermore, the Minkowski functionals can be generalized to
tensor-valued valuations. These Minkowski tensors quantify additional
morphometric information such as isotropy and homogeneity and thus
provide a comprehensive view of the available morphological
information. Incorporating this additional information may increase
the sensitivity of the morphometric analysis even more.
The inherent potential of our methods provides some exciting
perspectives for data analysis in VHE gamma-ray astronomy.

\begin{acknowledgements}
  We thank the German science foundation (DFG) for the grants
  ME1361/11 'Random Fields' and ME1361/12 'Tensor Valuations' awarded
  as part of the DFG-Forschergruppe ``Geometry and Physics of Spatial
  Random Systems''.
  We thank the H.E.S.S. Collaboration for providing the data.
  We thank our referee Dmitri Pogosyan for his advice.
\end{acknowledgements}


\bibliographystyle{aa}
\bibliography{morphometric_analysis_aa_555_A38_2013}

\end{document}